\documentclass{iopjournal}

\usepackage[english]{babel}
\usepackage[utf8]{inputenc}
\usepackage[colorinlistoftodos, color=green!40, prependcaption]{todonotes}
\usepackage{amsmath}
\usepackage{mathtools}
\usepackage{physics}
\usepackage{xcolor}
\usepackage{graphicx}
\usepackage{subcaption}
\usepackage{adjustbox}
\usepackage{placeins}
\usepackage{lipsum}
\usepackage{csquotes}
\usepackage{siunitx}
\usepackage{caption}
\usepackage{matlab-prettifier}
\usepackage{titlesec}

\newcommand\vk{\boldsymbol{k}}
\newcommand\vx{\boldsymbol{x}}
\newcommand\vi{\boldsymbol{i}}
\newcommand\vj{\boldsymbol{j}}
\newcommand\vn{\boldsymbol{n}}
\newcommand\vP{\boldsymbol{P}}
\newcommand\vp{\boldsymbol{p}}
\newcommand\vq{\boldsymbol{q}}

\definecolor{darkred}{rgb}{0.7, 0.1, 0.1}

\begin{document}

\articletype{\,}

\title{Probing metric fluctuations with the spin of a particle in a quantum simulation}

\renewcommand*{\thefootnote}{\fnsymbol{footnote}}

\author{Jiannis K. Pachos$^1$\orcid{0000-0002-9775-4436}, Patricio Salgado-Rebolledo$^{2,3}$\orcid{0000-0001-6832-6785} and Martine Schut$^{4}\footnotemark$\orcid{0000-0002-5585-4578}}

\affil{$^1$School of Physics and Astronomy, University of Leeds, Leeds LS2 9JT, United Kingdom}
\affil{$^2$Asia Pacific Center for Theoretical Physics (APCTP), Pohang, Gyeongbuk 37673, Korea}
\affil{$^3$Instituto de Ciencias Exactas y Naturales (ICEN), Universidad Arturo Prat, 1111346 Iquique, Chile}
\affil{$^4$Centre for Quantum Technologies, National University of Singapore, 3 Science Drive 2, 117543, Singapore}

\email{j.k.pachos@leeds.ac.uk, salgado.rebolledo@apctp.org, m.schut@nus.edu.sg}

\footnotetext{Corresponding author}

\renewcommand*{\thefootnote}{\arabic{footnote}}

\begin{abstract}
Exploring potential empirical manifestations of quantum gravity is a challenging pursuit. In this study, we utilise a lattice representation of a $(2+1)$D massive gravity toy model interacting with Dirac fermions that can support specific spacetime fluctuations. We focus on the evolution of the fermion's spin due to its coupling to spacetime fluctuations. To monitor these dynamics, a minimal model is required that comprises two bosonic modes describing spacetime geometry fluctuations coupled to the spin of the fermion. A possible emulation of this system involves encoding spin degrees of freedom in the electronic states of an atom coupled to a bimodal optical cavity that provides the two bosonic modes. Our proposal introduces a novel approach for modelling the effect of interactions between quantum gravity and matter that can be probed with current technology.
\end{abstract}

\vspace{5pt}
\section{Introduction}
\vspace{10pt}
\label{sec:intro}

Understanding the fundamental nature of spacetime at quantum scales is a central challenge in contemporary theoretical physics. Quantum gravity, the sought-after theory unifying general relativity and quantum physics, remains elusive, with several theoretical and empirical uncertainties. Apart from its conceptual difficulties, this theory is strongly interacting and is therefore often too complex to investigate analytically or numerically. Despite significant advances in theoretical frameworks such as lower-dimensional quantum gravity, string theory, and loop quantum gravity, observations and empirical validations of quantum gravitational effects remain well beyond the reach of current technologies \cite{Hossenfelder:2010zj,Carney:2018ofe}.

An interesting possibility for probing gravitational degrees of freedom and studying properties that are otherwise inaccessible is provided by condensed matter physics, where collective excitations that resemble gravitons have recently been identified. For example, massive gravitons have been shown to emerge in the $^3$He-B superfluid, where gravitational effects arise from fermionic bilinears following symmetry breaking \cite{Volovik:2021wut}. This bears a compelling resemblance to Diakonov's proposal for a lattice-regularized quantum gravity theory, which involves a massless emergent gravitational field \cite{Diakonov:2011im}.

Another significant area of exploration involves certain collective excitations in fractional quantum Hall liquids, which remarkably behave as analogue gravitons. Specifically, the Girvin-MacDonald-Platzman mode of the fractional quantum Hall effect \cite{PhysRevLett.54.581,PhysRevB.33.2481}, often termed the magnetoroton, is recognised as a massive graviton excitation in its long-wavelength limit \cite{PhysRevLett.117.216403,PhysRevX.7.041032,PhysRevLett.120.141601}. Moreover, the predicted properties of these excitations are in agreement with current experimental observations \cite{Pinczuk2024,PhysRevLett.70.3983,PhysRevLett.84.546}. However, the strong interactions inherent to fractional quantum Hall liquids make them analytically intractable, highlighting the need for alternative approaches that can emulate their emergent features.

To overcome these analytical limitations, an alternative strategy involves engineering systems capable of effectively simulating quantum signatures of gravity. Such quantum simulations offer the advantage of accessing a broad spectrum of coupling regimes. In particular, they enable the amplification of properties of gravity that are otherwise experimentally inaccessible due to gravity's very weak coupling to matter. Recent progress in simulating quantum gauge theories, for instance, has provided valuable insights into complex phenomena such as quark confinement \cite{zhou2021thermalization}. While various platforms exist for simulating classical gravity, such as deforming graphene-like systems to produce non-trivial extrinsic geometry \cite{Iorio:2017vtw,PhysRevD.101.036021} or creating tunnelling-coupling inhomogeneities that generate intrinsic geometries \cite{Boada:2010sh,PhysRevB.101.245116}, simulating quantum gravity has remained an open challenge. Roadblocks persist at both conceptual levels, such as realising quantum fluctuations of spacetime, and practical levels, such as engineering the intricate self-interactions characteristic of gravitational theories.

This work is motivated by the importance of bridging the gap between theory and experiment in quantum gravity. We propose an experimental setup that can simulate quantum metric fluctuations, thus facilitating the identification of novel physical manifestations of quantum gravity. Our starting point is based on a lattice representation of a particular model for metric perturbations in $(2+1)$ dimensions coupled to Dirac fermions \cite{Salgado-Rebolledo:2021zbx}. Unlike the model considered in \cite{Salgado-Rebolledo:2021zbx}, the fluctuations of the spatial geometry considered here are described by a traceless symmetric tensor, which can be associated with a unimodular spatial metric. Therefore, our lattice model gives rise to non-relativistic massive metric perturbations similar to the massive gravitons identified in the study of the Girvin--MacDonald--Platzman mode in the fractional quantum Hall effect \cite{PhysRevX.7.041032,PhysRevLett.120.141601}. This framework allows us to focus on a simplified set of spacetime fluctuations that still exhibits rich dynamical behaviour. We employ the interaction between matter and spacetime fluctuations to use the spin of the Dirac fermions as a probe of the gravitational dynamics.

The lattice representation of quantum gravity coupled to matter also has the advantage of being readily amenable to experimental simulation, thus bridging the gap between theory and experiment in this field of research. In particular, we identify the smallest instance of the lattice model that encodes the coupling between spacetime fluctuations and the spin of the Dirac fermion. We then propose an experimental setup for the quantum simulation of this minimal model. This setup involves encoding the spin degrees of freedom in the electronic levels of a single atom placed inside a bimodal optical cavity that encodes two bosonic modes responsible for the spacetime fluctuations. By varying the coupling between the atom and the cavity, we are able to use the atom as a probe to identify the dynamics of the simulated spacetime fluctuations. We identify dynamical signatures of the spin in the small-coupling regime, illustrating how spacetime fluctuations influence matter evolution at the quantum level. These findings open a pathway for experimental probes of quantum gravitational phenomena, possibly by enhancing the effect that fluctuating gravity can have on matter. Moreover, our research establishes a platform for simulating more intricate models of gravity--matter interactions in future work. Beyond their fundamental significance, such quantum simulations could shed light on information flow, entanglement dynamics, and decoherence in curved quantum geometries, offering valuable insights for both quantum gravity and quantum technologies.

The aim of this work is not to infer gravitational dynamics from experimental data, but to identify and characterise qualitative signatures that can arise when a quantum probe couples to quantised metric fluctuations within a controlled and experimentally simulatable setting. For this reason, we adopt a minimal toy model in which the gravitational sector is specified \emph{a priori}. The conclusions should therefore be understood within the scope of this modelling choice: different gravitational actions may lead to quantitatively different dynamics, and our focus is on robust mechanisms, such as entanglement generation, rather than on action-specific predictions.

While our work probes internal degrees of freedom (spin) coupled to quantised geometry, there are other proposals in the literature based on mass interferometry that probe spatial superpositions of heavy objects. Such experimental platforms for tests of the quantum nature of gravity typically consider interferometry and heavy masses ($\sim \mu\mathrm{g}$). These approaches focus either on gravity-induced entanglement between massive systems or on gravity-induced loss of coherence. For example, the QGEM proposal \cite{bose2017spin,MarlettoVedral} considers a table-top test consisting of two matter-wave interferometers that entangle solely through their mutual gravitational interaction. Several experimental platforms have been proposed to implement this idea, such as magnetically levitated nanoparticles, earth-based or space-based free-fall experiments with nanospheres, and mechanical resonators \cite{adelberger2022snowmass,kaltenbaek2012macroscopic}. A complementary line of investigation concerns gravitationally induced decoherence, which also requires long-lived quantum superpositions of heavy masses, among other requirements. There are proposals to create such superpositions in optical traps, electromagnetic traps, near-field interferometers, and related setups \cite{bassi2013models}. Finally, there are discussions about the possibility of testing quantum gravity predictions in gravitational-wave detectors \cite{barack2019black}.

\vspace{5pt}
\section{Lattice representation of Dirac field coupled to a massive spin-2 field}
\vspace{10pt}

We now consider a massless Dirac fermion field coupled to metric perturbations described by a toy-model Hamiltonian for a bosonic massive spin-2 field. In $2+1$ dimensions, pure Einstein gravity has no local propagating degrees of freedom, so its linearised (massless) limit does not provide dynamical metric fluctuations that could act as a nontrivial environment for a local matter probe. To obtain the simplest nontrivial dynamics while keeping the gravitational sector quadratic and analytically tractable, we therefore adopt a Fierz-Pauli massive spin-2 model and restrict to traceless spatial fluctuations, which yields two independent bosonic modes. This choice captures the minimal content required for time-dependent local metric fluctuations and is well suited to the reduction to a minimal spin-boson system used later. Subsequently, we consider a lattice model that describes this composite system in its low-energy limit.

\subsection{Carrollian massive spin-2 field}

For simplicity we consider purely spatial metric fluctuations $h_{ij}$ around a flat-space background. We adopt Cartesian coordinates $x^i=\{x,y\}$ so that the spatial metric tensor has the form
\begin{equation}\label{eqn:metric}
g_{ij}= \delta_{ij}+ h_{ij}.
\end{equation}
Moreover, we restrict $h_{ij}$ to be traceless, i.e.\ $h\equiv\delta^{ij}h_{ij}=0$. This has the consequence that the metric determinant is fixed to be ${\rm det}(g)=1$, a characteristic of unimodular gravity models. Since Einstein gravity in $2+1$ dimensions lacks propagating degrees of freedom \cite{Deser:1983tn}, the massless Fierz-Pauli action obtained by linearising the Einstein-Hilbert Lagrangian has trivial dynamics for $h_{ij}$. To introduce non-trivial dynamics, we instead model $h_{ij}$ as a massive spin-2 field. The adopted Fierz--Pauli action for a relativistic massive spin-2 field in $2+1$ dimensions describes two local propagating degrees of freedom \cite{Bergshoeff:2009fj}, thus having behaviour very similar to that of linearised Einstein gravity in four dimensions. For our purposes, it is sufficient to focus on the temporal dynamics of the metric fluctuations, neglecting spatial derivatives of $h_{ij}$ while still capturing the essential physical behaviour and complexity of the system. We therefore consider the following action principle
\begin{equation}
\label{Sgrav1}
S_{\rm gr}= \frac{1}{64\pi G} \int \text{d}t \int \text{d}^2 x\; \big[\dot h_{ij}\dot h^{ij} -\dot h^2 
-\mu^2\left( h_{ij} h^{ij} -h^2 \right)+\xi \,h\big],
\end{equation}
where $\mu$ is a mass parameter and $\xi$ is a Lagrange multiplier enforcing the tracelessness condition. This non-relativistic Hamiltonian for metric perturbations can be obtained as the electric-type Carrollian limit of the Fierz--Pauli Lagrangian for a spin-2 field \cite{Henneaux:2021yzg} after integrating out the field components $h_{0i}$ and by identifying $h_{00}=\xi$. By rescaling the metric fluctuations as
\begin{equation}\label{rescalingh}
h_{ij}\rightarrow \sqrt{16 \pi G} \, h_{ij},
\end{equation}
the canonical momenta of the model are given by
\begin{equation}
p_{ij}=\frac12\left(
\dot h_{ij}-\dot h \delta_{ij}
\right), \qquad p_\xi =0,
\end{equation}
satisfying canonical Poisson brackets
\begin{equation}
\{ h_{ij}(\vx), p^{kl}(\vx') \}
= 
\left(  \delta_i^k \delta_j^l + \delta_i^l \delta_j^k 
\right)\delta^{(2)}(\vx - \vx'), \qquad \{\xi(\vx),p_\xi(\vx')\}=\delta^{(2)}(\vx - \vx').
\end{equation}
Action~\eqref{Sgrav1} defines a constrained Hamiltonian system. The momentum conjugate to the Lagrange multiplier $\xi$ vanishes identically, $p_\xi=0$, generating a primary constraint whose preservation under time evolution yields secondary constraints enforcing the tracelessness of $h_{ij}$ and of its conjugate momentum. Following the standard Dirac procedure, these second-class constraints $\phi_1=p_\xi$, $\phi_2=h$, $\phi_3=p$, and $\phi_4=\xi$ (with $p\equiv\delta^{ij}p_{ij}$) are eliminated by introducing Dirac brackets, resulting in a reduced phase space with two independent dynamical degrees of freedom and the Hamiltonian
\begin{equation}\label{Hgrav1}
 H_{\rm gr}=  \int  \text{d}^2 x\; \left(p_{ij}p^{ij} +\frac{\mu^2}{4} h_{ij} h^{ij}\right),
\end{equation}
together with the Dirac bracket
\begin{equation}
\{ h_{ij}(\vx), p^{kl}(\vx') \}^*
= 
\left(  \delta_i^k \delta_j^l + \delta_i^l \delta_j^k 
- \delta_{ij} \delta^{kl} \right)\delta^{(2)}(\vx - \vx').
\end{equation}
To quantise the model, $h_{ij}$ and $p_{ij}$ are promoted to operators with commutation relations compatible with the Dirac bracket $\{\,,\,\}^*\rightarrow -i[\,,\,]$. Equivalently, we parametrise the independent components of $h_{ij}$ and $p_{ij}$ in terms of operators $\tilde\alpha$ and $\tilde\beta$ satisfying the commutation relations
\begin{equation}\label{commops}
[\tilde\alpha(\vx),\tilde\alpha^\dagger(\vx')]= \delta^{(2)}(\vx -\vx')=
[\tilde\beta(\vx),\tilde\beta^\dagger(\vx')], 
\end{equation}
as
\begin{equation}\label{alphabeta}
\begin{aligned}
& h_{11}(\vx)=\frac1{\sqrt2} \left( \tilde \alpha(\vx)+  \tilde \alpha^\dagger(\vx) \right), \qquad h_{12}(\vx)=\frac1{\sqrt2}\left( \tilde \beta(\vx)+  \tilde \beta^\dagger(\vx)\right), \\
&p_{11}(\vx)=\frac i{\sqrt2}\left(  \tilde \alpha(\vx)-  \tilde \alpha^ \dagger(\vx)\right), \qquad p_{12}(\vx)=\frac i{\sqrt2}\left(  \tilde \beta(\vx)-  \tilde \beta^ \dagger(\vx)\right).
\end{aligned}
\end{equation}
In momentum space the Hamiltonian can be written as $H_{\rm gr}= H[\tilde\alpha] +H[\tilde\beta]$ with
\begin{equation}\label{Halphacontinuum}
H[\tilde\alpha]=2\int \text{d}^2 k\bigg[
\left( \frac{\mu^2}4-1\right)\tilde\alpha(\vk)\tilde \alpha(-\vk)
+\left( \frac{\mu^2}4+1\right)
\tilde \alpha(\vk)\tilde \alpha^\dagger(\vk) + {\rm h.c.}
\bigg],
\end{equation}
and similarly for $H[\tilde\beta]$. By applying the Bogoliubov transformation
\begin{equation}\label{BTransformation}
\tilde \alpha(\vk) = \alpha(\vk) \cosh{r} - \alpha^\dagger(-\vk) \sinh r,\,\,\,\,\,\,\,
\tilde \beta(\vk) = \beta(\vk) \cosh r - \beta^\dagger(-\vk) \sinh r, 
\end{equation}
with
\begin{equation}\label{defcoshsinh}
\cosh{2r}=\frac{\mu}4 + \frac{1}{\mu},\qquad
\sinh{2r}=\frac{\mu}4 - \frac{1}{\mu},
\end{equation}
the Hamiltonian takes the form
\begin{equation}\label{Hgrav3}
\begin{aligned}
H_{\rm gr}&= 2\mu\,\int \text{d}^2 k \bigg[
\alpha^\dagger(\vk) \alpha(\vk)+ \beta^\dagger(\vk) \beta(\vk) +\frac12\delta(0) \bigg],
\end{aligned}
\end{equation}
where the final constant term can be dropped.

\subsection{Coupling to Dirac fermion}

To describe the coupling of these fluctuations to a Dirac field, we introduce the spatial zweibein $e^a_i$, with $a=1,2$, such that $g_{ij}=\delta_{ab} e^a_i e^b_j$. A possible choice leading to the spatial metric \eqref{eqn:metric}, after taking into account the rescaling \eqref{rescalingh}, is
\begin{equation}\label{zweibein}
e^a_i = \delta^a_i+ 2\sqrt{\pi G}\delta^{aj} h_{ij}.
\end{equation}
The Dirac Hamiltonian on this geometry is given by
\begin{equation}
H_{\rm Dirac}=\frac{i}2\left(\bar\psi e^i_a\gamma^a \partial_i \psi - \partial_i \bar\psi e^i_a\gamma^a \psi\right),
\end{equation}
where $\gamma^a$ are gamma matrices and $e^i_a$ is the inverse zweibein, satisfying $e^i_a e^a_j=\delta^i_j$ and $e^i_a e^b_i=\delta^b_a$.

The Hamiltonian of the full model is then given by 
\begin{equation}
\label{GQFThamiltonian}
H= \int  \text{d}^2 x\; \psi^\dagger \; h\; \psi  +  H_\text{gr},
\end{equation}
where the single-particle Dirac Hamiltonian $h$ reads
\begin{equation}
\label{singpart}
h=\gamma^0 \left( \delta^i_a\gamma^a - 2\sqrt{\pi G} \delta_{aj} h^{ij}\right)(-i \partial_i)  +\frac{i\sqrt{\pi G}}2\delta_{aj}
 \partial_i h^{ij}  \gamma^0  \gamma^a ,
\end{equation}
describing $(2+1)$D Dirac fermions coupled to semiclassical gravity fluctuations. In the following we present a lattice model that gives rise to this Hamiltonian in the continuum limit.

\subsection{Lattice Hamiltonian}

\begin{figure}[t]
    \centering
\includegraphics[width=0.55\linewidth]{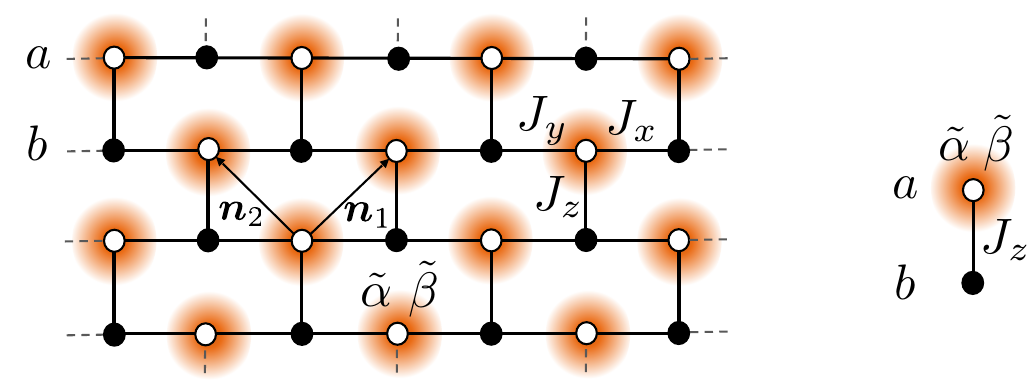}
\caption{(Left) The brick-wall lattice representation of the $(2+1)$D Dirac fermions coupled to spacetime fluctuations given by Hamiltonian \eqref{OLH}. The unit cell includes two fermionic modes, $a$ and $b$, that encode the spin of the Dirac field. The bosonic modes $\tilde\alpha$ and $\tilde\beta$ describe spacetime fluctuations and are coupled to the fermions through their tunnelling. The basis vectors $\boldsymbol{n}_1$ and $\boldsymbol{n}_2$ span the brick-wall lattice of fermions and $J_x$, $J_y$, and $J_z$ are the fermionic tunnelling couplings that depend on the bosonic modes $\tilde \alpha$ and $\tilde\beta$. (Right) The minimal system considered here, with two fermionic modes that encode a single spin and two bosonic modes coupled to the fermions.}
\label{fig:Figure1}
\end{figure}

The $(2+1)$D Dirac fermion coupled to the massive spin-2 field can effectively arise from a two-dimensional lattice model. In \cite{Salgado-Rebolledo:2021zbx}, a two-dimensional lattice of interacting bosonic and fermionic modes is presented that gives rise to a Hamiltonian resembling \eqref{GQFThamiltonian} in its low-energy limit. A similar strategy is adopted here, where we start with a brick-wall lattice of fermions, as shown in Fig.~\ref{fig:Figure1} (left). The lattice Hamiltonian $H_{\rm latt}$ is given by
\begin{equation}\label{OLH}
H_{\rm latt}=H_{\text{f-b}}+ H_\text{b},
\end{equation}
where the fermionic and bosonic term $H_{\text{f-b}}$ gives rise to Dirac fermions coupled to gravity fluctuations, while the bosonic term $H_\text{b}$ corresponds to the pure gravity sector. The gravitational lattice Hamiltonian is given by
\begin{equation}
H_{\rm b}= H[\tilde\alpha] +H[\tilde\beta],
\end{equation}
where
\begin{equation}\label{Halpha}
H[\tilde\alpha] =
\sum_{\vi}\bigg( 
\frac{\mu^2}2 (\tilde\alpha_{\vi}+\tilde\alpha^\dagger_{\vi}) (
\tilde\alpha_{\vi}+\tilde\alpha^\dagger_{\vi})
-2 (\tilde\alpha_{\vi}-\tilde\alpha^\dagger_{\vi}) (
\tilde\alpha_{\vi}-\tilde\alpha^\dagger_{\vi})
\bigg).
\end{equation}
The fermion-boson interaction term in \eqref{OLH} describes a two-dimensional brick-wall lattice, where fermionic operators $a$ and $b$ live at its vertices and the bosons couple to them when the fermions tunnel to neighbouring sites \cite{Salgado-Rebolledo:2021zbx}. These fermions satisfy 
\begin{equation}
[a^\dagger_{\vi},a_{\vj}]=2\delta_{\vi,\vj}=
[b^\dagger_{\vi},b_{\vj}],
\end{equation}
and are subject to the tunnelling Hamiltonian
\begin{equation}\label{FHlattice}
H_{\text{f-b}}= \sum_{\vi} \bigg( a^\dagger_{\vi} J_X(\vi) b_{\vi+\vn_1} + a^\dagger_{\vi} J_Y(\vi) b_{\vi+\vn_2} 
+ a^\dagger_{\vi} J_Z(\vi) b_{\vi}\bigg) + {\rm h.c.},
\end{equation}
where $\vi=(i_x,i_y)$ gives the position of unit cells on the lattice and the vectors
\begin{equation}\label{nvectors}
\vn_1=\left(-\frac{1}{\sqrt2},\frac{1}{\sqrt2}\right),\qquad
\vn_2=\left(\frac{1}{\sqrt2},\frac{1}{\sqrt2}\right),
\end{equation}
transport between unit cells (see Fig.~\ref{fig:Figure1}). When linearised around its Fermi points, this type of Hamiltonian describes the coupling of a fermion field to a background metric whose components are encoded in the tunnelling couplings $J_\alpha(\boldsymbol i)$, $\alpha=X,Y,Z$ \cite{PhysRevB.101.245116}. Therefore, to couple the fermions to dynamical gravitational fluctuations, we consider tunnelling couplings $J_\alpha(\boldsymbol i)$ given by a constant background value together with a background perturbation described precisely by the bosonic fluctuations $\tilde \alpha_{\vi}$ and $\tilde \beta_{\vi}$:
\begin{equation}\label{JXJYJZ}
J_X(\vi)=J_Y(\vi)= \frac{1}{\sqrt2} J_Z(\vi) 
= 1
+ i \sqrt{2\pi G}(\tilde\alpha_{\vi}+\tilde\alpha^\dagger_{\vi})
-\sqrt{2\pi G}(\tilde\beta_{\vi}+\tilde\beta^\dagger_{\vi}).
\end{equation}
Following a similar procedure to that in \cite{Salgado-Rebolledo:2021zbx}, one can show that the lattice of tunnelling fermions and bosons shown in Fig.~\ref{fig:Figure1} (left) gives rise, in its low-energy limit, to a $(2+1)$D Dirac fermion coupled to massive metric fluctuations, as given in \eqref{GQFThamiltonian}. While it is straightforward to see that the continuum limit of \eqref{Halpha} in momentum space leads to Eq.~\eqref{Halphacontinuum}, the fermionic Hamiltonian requires some extra analysis. First, we consider the free Hamiltonian by setting $\tilde\alpha_{\vi}=0=\tilde\beta_{\vi}$ in \eqref{FHlattice}, i.e.
\begin{equation}
J_X=J_Y=J_Z/\sqrt2=1,
\end{equation}
and passing to momentum space, which leads to
\begin{equation}
H^{(free)}_{\text{f-b}}= \sum_{\vp} f(\vp)\;a^\dagger_{\vp} b_{\vp} + {\rm h.c.},\qquad f(\vp)=J_X e^{i\vp\cdot\vn_1}+J_Y e^{i\vp\cdot\vn_2}+J_Z .
\end{equation}
From this expression one can find the Fermi points of the free Hamiltonian $\vp=\vP$ such that $f(\vP)=0$, which yields
\begin{equation}
\vP_{\pm}= \mp \left(\frac{\pi}{2\sqrt2}, \sqrt2\pi \right).
\end{equation}
Second, restoring the perturbations and expanding the Hamiltonian \eqref{FHlattice} in momentum space around these Fermi points, one finds
\begin{equation}\label{H_lattsingle}
H_{\text{f-b}}=\sum_{\vp,\vq}\Psi^{+\dagger}_{\vp} \,h^+(\vp,\vq)\,\Psi^+_{\vq}
+ 
\sum_{\vp,\vq}\Psi^{-\dagger}_{\vp} \,h^-(\vp,\vq)\,\Psi^-_{\vq},
\end{equation}
where we have defined
\begin{equation}
\Psi^\pm_{\vp}=\begin{pmatrix}
a^\pm_{\vp}\\
b^\pm_{\vp}
\end{pmatrix}
,\qquad h^\pm(\vp,\vq)=\mp A_{\vp-\vq}\sigma^x q_x \mp C_{\vp-\vq}\sigma^x q_y - B_{\vp-\vq}\sigma^y q_y -D_{\vp-\vq}\sigma^y q_x,
\end{equation}
with $\sigma^x$ and $\sigma^y$ being Pauli matrices, and
\begin{equation}
A_{\vk}=1-\sqrt{2\pi G} (\tilde\alpha_{\vk}+\tilde\alpha_{-\vk}^\dagger),
\quad 
B_{\vk}= 1+\sqrt{2\pi G} (\tilde\alpha_{\vk}+\tilde\alpha_{-\vk}^\dagger),
\quad
C_{\vk}=D_{\vk}=-\sqrt{2\pi G} (\tilde\beta_{\vk}+\tilde\beta_{-\vk}^\dagger).
\end{equation}
Lastly, returning to position space and defining the four-fermion field $\psi$ and the gamma matrices $\gamma^0$ and $\gamma^i$ as
\begin{equation}
\psi=
\begin{pmatrix} 
a^+\\
b^+\\
b^-\\
a^-
\end{pmatrix}
,\quad
\gamma^0=\begin{pmatrix} 
 0 &-\boldsymbol 1\\
 \boldsymbol1 &0 
\end{pmatrix},\quad
\gamma^i=\begin{pmatrix} 
 0 &\sigma^i \\
\sigma^i &0 
\end{pmatrix},
\end{equation}
allows one to write $H_{\text{f-b}}$ in the Dirac form $\int  \text{d}^2 x\; \psi^\dagger h\, \psi$. When expressing the modes $\tilde\alpha$ and $\tilde\beta$ in terms of the metric perturbations as given in Eq.~\eqref{alphabeta}, the single-particle Hamiltonian $h$ matches precisely \eqref{singpart}. Note that although Eq.~\eqref{H_lattsingle} couples fermionic modes with different momenta, this does not imply spatial non-locality. It is simply the momentum-space representation of the local lattice Hamiltonian in Eq.~\eqref{FHlattice}, with locality restored upon transforming back to position space.

\vspace{5pt}
\section{Dynamical evolution of a spin-$1/2$ due to geometry fluctuations}
\vspace{10pt}

In the previous section we presented the lattice Hamiltonian that effectively encodes the dynamics of a Dirac field coupled to a fluctuating geometric background. A proposal for how to realise a similar lattice model with optical lattices and a mixture of bosonic and fermionic atoms was given in \cite{Salgado-Rebolledo:2021zbx}. Here we consider only the evolution of the Dirac fermion's spin due to the fluctuating background. The minimal subsystem of the lattice required to encode and monitor this evolution comprises two fermionic modes that encode the spin of the particle and two bosonic eigenmodes that couple resonantly to the fermionic ones. Such a minimal system provides an analytically tractable way to predict the time evolution of the system. Moreover, it lends itself to quantum simulation with optical cavity technology, as we shall see in this section.

\vspace{5pt}
\subsection{Spin-$1/2$ coupled to quantum geometry fluctuations}
\vspace{10pt}

We can isolate the dynamics that the spin-2 field induces on the spin-$1/2$ of the Dirac particle by considering a single fermionic unit cell with $\vi={\mathbf 0}$. The restriction to a single unit cell corresponds physically to a tightly confined Dirac particle whose spatial motion is frozen, so that only its internal spin degree of freedom remains dynamical while it couples to the full set of resonant bosonic (gravitational) modes. Omitting the site label, the fermion-boson Hamiltonian reduces to
\begin{equation}\label{Hfbsimplified}
\begin{aligned}
H_{\text{f-b}} =
 J_Z a^\dagger b + J_Z^\dagger b^\dagger a,
\end{aligned}
\end{equation}
where the tunnelling coupling $J_z$ is controlled by the bosonic fields as
\begin{equation}
J_Z 
= \sqrt2
+2 i \sqrt{\pi G}(\tilde\alpha+\tilde\alpha^\dagger)
-2\sqrt{\pi G}(\tilde\beta+\tilde\beta^\dagger).
\end{equation}
We construct a minimal model by considering the Hamiltonian \eqref{Hfbsimplified} together with the Hamiltonian for gravitational fluctuations in momentum space in the continuum \eqref{Hgrav3}, thus emulating a single spin degree of freedom. This implies that the mode $\tilde \alpha =\tilde \alpha_{\vi={\mathbf 0}}$ is Fourier transformed as
\begin{equation}
\tilde \alpha=  \int \frac{\text{d}^2 k}{2\pi} \tilde \alpha(\vk)
\end{equation}
and similarly for $\tilde \beta$. Using relation \eqref{BTransformation}, which expresses $\tilde \alpha$ and $\tilde \beta$ in terms of the eigenmodes $\alpha$ and $\beta$, the full Hamiltonian of the system takes the form
\begin{equation}
\begin{aligned}
H= \sqrt2 \bigg(1  -\sqrt{\frac{G}{\mu\pi}}\int \text{d}^2 k \big(\beta(\vk)+&\beta^\dagger(\vk)
- i \alpha(\vk)- i \alpha^\dagger(\vk)\big)
\bigg)a^\dagger_{\mathbf 0} b_{\mathbf 0}
\\&
+ \mu\,\int \text{d}^2 k\bigg[
 \alpha^\dagger(\vk) \alpha(\vk) + \beta^\dagger(\vk) \beta(\vk)\bigg]
  + {\rm h.c.}.
\end{aligned}
\end{equation}
Adopting polar coordinates $k_x=k\cos\theta$ and $k_y=k\sin\theta$, and expanding the bosonic operators $\alpha$ and $\beta$ in Fourier modes, $\alpha(\vk) = \sum_ n \alpha_n(k) e^{in\theta}$, leads to
\begin{equation}
H=  \sqrt2 a^\dagger b +\int_0^\infty k\text{d}k 
\Bigg[
2\pi\mu \left(\alpha^\dagger_0\alpha_0+\beta^\dagger_0 \beta_0\right)
-2\sqrt{\frac{\pi G}{\mu}}\big(\beta_0+\beta^\dagger_0- i \alpha_0- i \alpha^\dagger_0\big)a^\dagger b \Bigg]
+ {\rm h.c.},
\end{equation}
where we have neglected the Fourier modes $\alpha_m(k)$ and $\beta_m(k)$ for $m\neq0$ since they decouple from the fermionic modes. Out of all possible momenta, the dominant terms in this Hamiltonian are those for which the resonant condition $\sqrt2\approx2\pi k \mu$ holds, which defines a circle in momentum space of radius $k_R=\frac{1}{\sqrt2 \pi\mu}$. Dropping the operator labels for the bosonic modes, the Hamiltonian can be rewritten as
\begin{equation}
\label{eq:H-final}
H=\sqrt2 \sigma^x + \sqrt2 \left(\alpha^\dagger \alpha + \beta^\dagger \beta \right)
+g_{\text{s-b}}\left[ \left(\beta+\beta^\dagger\right)\sigma^x +\left(\alpha+\alpha^\dagger\right)\sigma^y\right],
\end{equation}
where $\sigma^x=a^\dagger b+b^\dagger a$ and
$\sigma^y=-i\left(a^\dagger b-b^\dagger a\right)$ perform the rotation of a spin encoded on the two sites $a$ and $b$ of the unit cell ${\boldsymbol i}$, as shown in Fig.~\ref{fig:Figure1}. We also define the effective spin-boson coupling
\begin{equation}
g_{\text{s-b}}=-\frac{\sqrt{2 G}}{\sqrt \pi \mu^{3/2}}.
\end{equation}

Note that the resonance condition makes the self-interaction strengths of the bosonic and fermionic parts equal. This approximation is valid when the effective spin--boson coupling $g_\text{s-b}$ is small compared with the bosonic frequency scale, i.e.\ in the weak-coupling regime $G\ll \mu$. In this regime, the dynamics is dominated by the near-resonant modes and the model reduces effectively to a few-mode spin--boson system. For larger values of $G$, off-resonant modes of the continuum contribute significantly to the dynamics and the two-mode truncation is no longer quantitatively reliable. In the following we therefore focus our physical interpretation on the weak-coupling regime, where the approximation is justified. This minimal system of a spin-$1/2$ particle and two bosonic modes offers an ideal platform for investigating the dynamical evolution of a particle's spin when coupled to a gravitational field in the weak-coupling regime.

\vspace{5pt}
\subsection{Dynamical evolution of spin}
\vspace{10pt}
\label{sec:dynamics}

Hamiltonian~\eqref{eq:H-final} determines the evolution of a particle's spin due to its coupling with quantum fluctuations of the gravitational field. The discussion below focuses on the small-coupling limit $G\ll\mu$, where the time evolution under Hamiltonian~\eqref{eq:H-final} shows coherent oscillations of spin rotations and fluctuations in the bosonic populations.

The spin state, $|{\rm spin}\rangle$, can be initially prepared along the $x$, $y$, or $z$ directions, and the corresponding time evolutions are considered separately below. The graviton modes are prepared in their ground state, which for the bosonic part of \eqref{eq:H-final} corresponds to the zero-population Fock state $\ket{0}_\alpha \otimes \ket{0}_\beta$. Hence, the initial state of the system is given by
\begin{equation}\label{eq:initial-state}
 \ket{\psi_0} = \ket{\rm spin}_f\ket{0}_\alpha \ket{0}_\beta \, ,
\end{equation}
where the subscript $f$ denotes the fermionic state and the subscripts $\alpha$ and $\beta$ denote the two bosonic modes. The time evolution of this state,
\begin{equation}
\ket{\psi(t)} = e^{-i {H} t} \ket{\psi_0},
\end{equation}
under Hamiltonian~\eqref{eq:H-final} will in general produce oscillatory dynamics between the spin and the bosons. Let us first consider the case where $g_{\text{s-b}}=0$. Since the bosonic modes start in the ground state and are decoupled from the spin, there is no dynamics in the bosonic populations. If the spin is also prepared in an eigenstate of $\sigma^x$, then no evolution occurs, whereas any other initial spin state causes it to rotate due to the gravitational background classical field.

Now let us consider the case where $g_{\text{s-b}} \neq 0$. Even if the spin is prepared in the $x$-direction, it evolves because it is coupled non-trivially to the bosonic modes. Note that the evolutions corresponding to other initial spin states are asymmetric due to the $\sigma^x$ self-energy of the spin. Due to the coupling, the spin-$x$ population is exchanged with the $\beta$-mode population and its oscillation amplitude decays over time.

To monitor the time-dependent populations of the bosonic modes and the fermionic state, we define
\begin{equation}
n_\alpha (t)  = \bra{\psi(t)}\alpha^\dagger \alpha\ket{\psi(t)}, \qq{} n_\beta (t)  = \bra{\psi(t)}\beta^\dagger \beta\ket{\psi(t)}, \qq{}
\sigma_i(t) = \bra{\psi(t)}\hat{\sigma}^i\ket{\psi(t)},
\end{equation}
where $i=x,y,z$ for the Pauli spin operators.

In the following we numerically investigate the time evolution for various coupling strengths. Note that $g_{\text{s-b}}$ depends on $G$ and $\mu$. For simplicity we set $\mu=1$, while $G$ is restricted to the regime $G\ll1$, i.e.\ weak coupling. In the numerical simulation we must introduce a cutoff for the otherwise infinite-dimensional bosonic matrices $\alpha$ and $\beta$. Our numerical analysis shows that increasing the matrix size from $N=14$ to $N=15$ changes the bosonic populations by at most $0.5\%$. We therefore consider $N=14$ to be a good approximation for these bosonic matrices in our simulation. We note that this truncation does not modify the gravitational model or discretise the metric degrees of freedom. The continuum field theory is fixed from the outset, and only the infinite-dimensional Fock-space representation of the bosonic modes is truncated as a numerical approximation, whose validity is checked by convergence with increasing cutoff.

\subsubsection{Preparation in the spin-$x$ direction}

\begin{figure*}[bpt!]
\captionsetup[subfigure]{justification=centering}
    \centering 
        \begin{subfigure}[t]{0.48\textwidth}
        \centering
        \includegraphics[width=\linewidth]{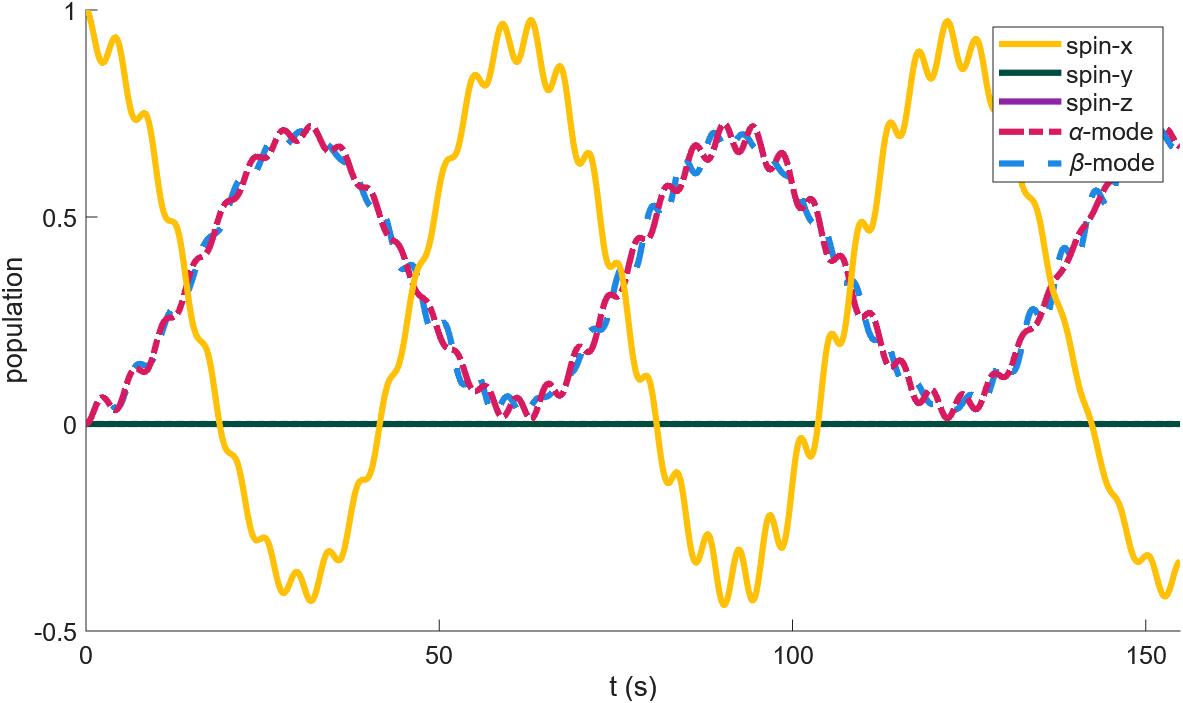}
        \caption{}
        \label{subfig:2D}
    \end{subfigure}%
    \hfill
    \begin{subfigure}[t]{0.48\textwidth}
        \centering
        \includegraphics[width=\linewidth]{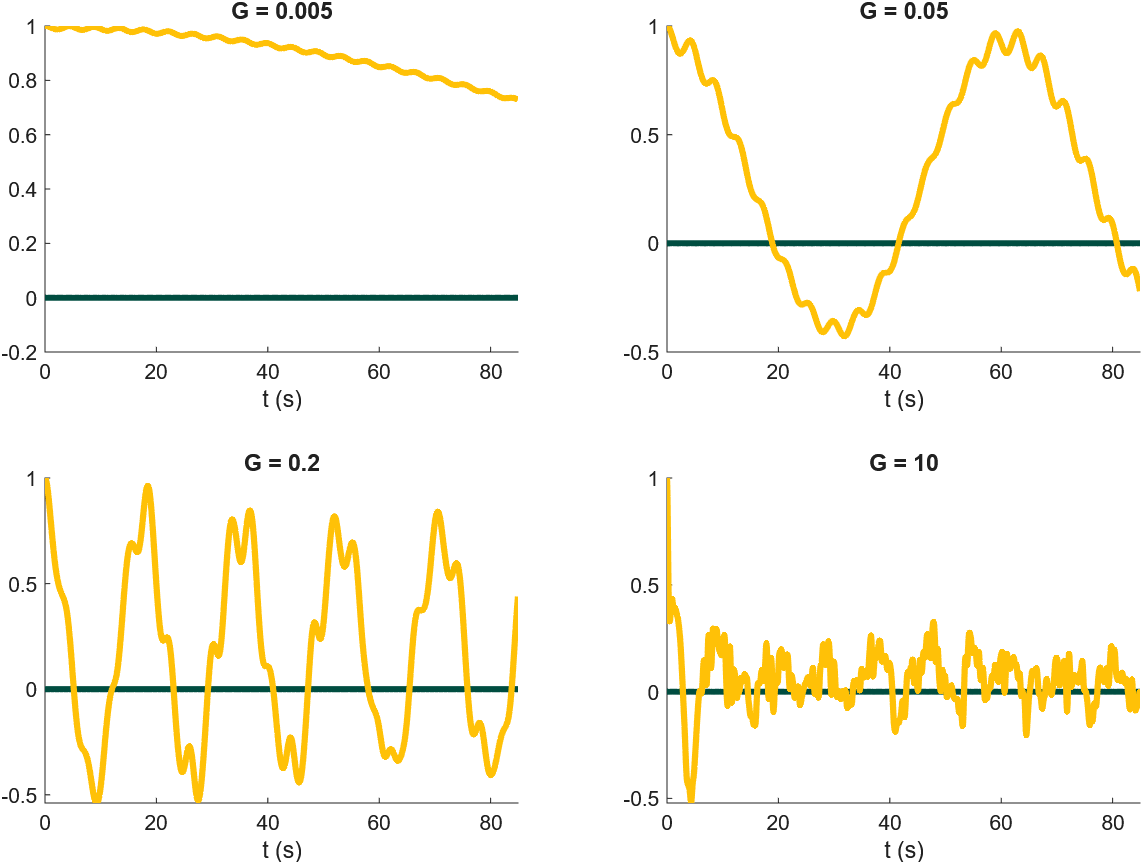}
        \caption{}
        \label{subfig:panel}
    \end{subfigure}
    \caption{Population dynamics for the system initialised with spin aligned along the $x$-axis ($\mu = 1$; see Hamiltonian~\eqref{eq:H-final}). (a) Time evolution of the spin and bosonic populations for $G = 0.05$, where the spin--boson interaction is much weaker than the bosonic self-interaction. The spin is initially prepared along the $x$-axis. The spin-$x$ population (solid) exhibits coherent exchange with the bosonic modes (dashed/dotted), while spin-$y$ and spin-$z$ remain unpopulated. (b) Spin-$x$ dynamics for increasing coupling strength $G$ (spin-$y$ is shown but does not have dynamics). Stronger coupling leads to faster and less smooth oscillations in the spin dynamics. The last panel shows $G\gg\mu$.}
    \label{fig:spin-x}
\end{figure*}

To study the interplay between spin and bosonic modes, we initialise the system with the spin pointing along the $x$-axis and track its evolution under Hamiltonian~\eqref{eq:H-final}. In the weak-coupling regime, where the spin--boson coupling is much smaller than the bosonic self-interaction strength, the spin-$x$ population exhibits regular, quasi-coherent oscillations, as shown in Fig.~\ref{subfig:2D}. These oscillations reflect the spin-boson interaction terms in \eqref{eq:H-final}: the coupling to the $\beta$ mode effectively adds a linear shift, which can be removed by displacing the operator $\beta$, while the coupling of spin-$x$ to the $\alpha$ mode is accompanied by a spin rotation via the $\sigma^y$ operator and resembles quantum Rabi oscillations in the weak-coupling limit and near resonance (see Appendix~\ref{app:JC}). Thus, there is a reversible exchange of excitation between the spin and the $\alpha$ bosonic mode, together with a displacement to a coherent state for the $\beta$ bosonic mode. Although the mechanisms behind the population dynamics differ for the two bosonic modes, they are both expected to scale as $\propto\sin^2$ (see Appendix~\ref{app:JC} and \cite{koch2023quantum}), which explains the oscillatory behaviour of the bosonic populations at small couplings. The frequency of the oscillations is controlled by the coupling $g_{\text{s-b}}$. The spin-$y$ and spin-$z$ populations remain unpopulated due to the initial spin alignment and the decoupled dynamics.

The small high-frequency fluctuations visible on top of the main oscillations in Fig.~\ref{subfig:2D} are due to the self-interactions of the bosons and the spin, with coupling $\sqrt{2}$, in \eqref{eq:H-final}. Figure~\ref{subfig:panel} illustrates how the spin-$x$ population evolves as a function of time for several couplings $G$. As the coupling strength $G$ increases, the micro-oscillations become more pronounced and disrupt the coherent oscillation pattern. For $G \ll \mu$, where the self-interaction dominates, the spin-$x$ population exhibits near-perfect revivals, indicating weak backaction from the bosonic environment. As $G$ increases, two effects become prominent: (i) the emergence of fast oscillations superimposed on the slower envelope and (ii) a decay in the amplitude of the envelope itself, signalling entanglement between the spin and the bosonic gravitational modes. For larger couplings $G>\mu$, the truncated model produces increasingly irregular dynamics, as shown in the last panel of Fig.~\ref{subfig:panel}. However, in this regime, contributions from the continuum of bosonic modes neglected in the minimal model become important. Therefore, the behaviour observed in the strong-coupling region should be regarded only as indicative of the breakdown of the few-mode approximation rather than as a physical prediction of the underlying gravity--matter model. Note that at $G=\pi$, the spin--boson coupling matches the bosonic self-interaction ($g_{\text{s-b}} = \sqrt{2}$), providing a representative scale at which the spin--boson coupling becomes comparable to the bosonic self-interaction in the truncated model.

\subsubsection{Preparation in the spin-$y$ or spin-$z$ direction}

\begin{figure*}[bpt]
\captionsetup[subfigure]{justification=centering}
    \centering 
    \begin{subfigure}[t]{0.47\textwidth}
        \centering
        \includegraphics[width=\linewidth]{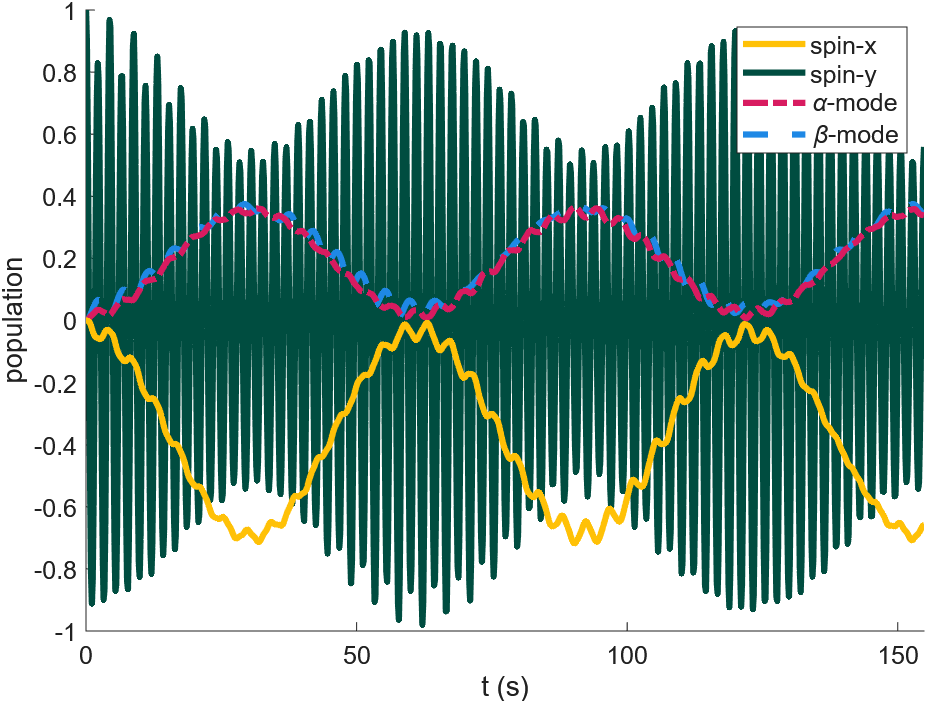}
        \caption{}
        \label{subfig:2D-y}
    \end{subfigure}%
    \hfill
    \begin{subfigure}[t]{0.47\textwidth}
        \centering
        \includegraphics[width=\linewidth]{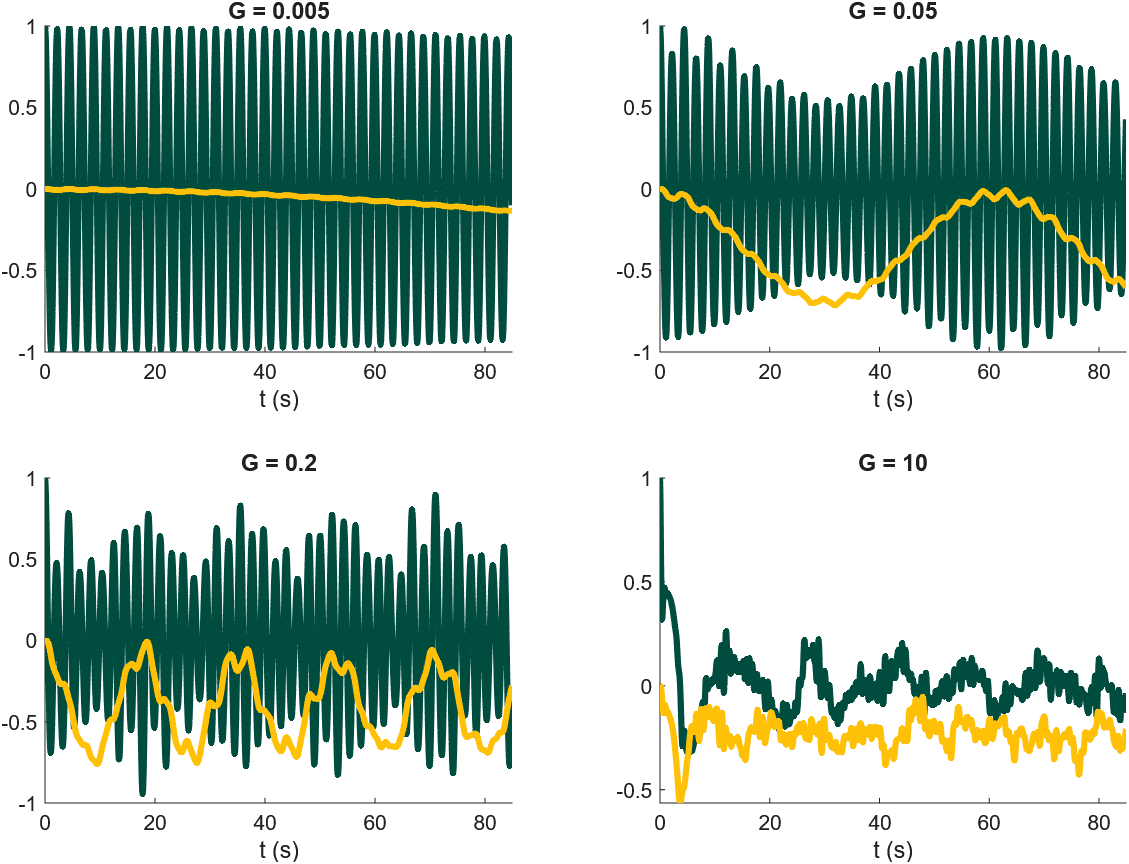}
        \caption{}
        \label{subfig:panel-y}
    \end{subfigure}
    \caption{Spin and bosonic population dynamics for a spin initially aligned along the $y$-axis ($\mu = 1$; see Hamiltonian~\eqref{eq:H-final}). The spin-$z$ population is not shown but closely follows spin-$y$ with a $\pi/2$ phase shift. (a) Population dynamics for $G = 0.05$. The spin-$y$ and spin-$z$ populations exhibit rapid oscillations with slowly varying envelopes. The spin-$x$ grows from zero and mediates coupling to the bosonic modes, which remain nearly identical. (b) Time evolution of the spin-$x$ (yellow line) and spin-$y$ (green envelope) populations for different coupling strengths $G$. As $G$ increases, coherence is gradually lost and the system transitions to increasingly irregular dynamics. At $G = \pi$, the spin--boson coupling equals the bosonic self-interaction strength.}
    \label{fig:spin-y}
\end{figure*}

The cases in which the spin is initially aligned along the $y$- or $z$-axis are similar up to a relabelling of spin components. Consequently, we present results only for the case of initial preparation along the $y$-direction. Figure~\ref{fig:spin-y} displays the time evolution of the system for various coupling strengths $G$, focusing on the spin and bosonic population dynamics.

Figure~\ref{subfig:2D-y} shows the full population dynamics for weak coupling $G = 0.05$. Initially, the spin is fully aligned along the $y$-axis, with zero population in spin-$x$ and spin-$z$. As time evolves, rapid oscillations develop between the spin-$y$ and spin-$z$ populations, reflecting precession in the $y$--$z$ plane. The envelope of this oscillation exhibits slower modulations, corresponding to the gradual exchange of population with and/or displacement of the bosonic modes. Meanwhile, the spin-$x$ population grows from zero and tracks this envelope, acting as a mediator between the spin subspace and the bosonic environment. The bosonic modes themselves remain nearly identical throughout the evolution due to the symmetry of their coupling, despite the different mechanisms behind their population evolution.

To highlight how the spin coherence degrades with increasing coupling strength, Fig.~\ref{subfig:panel-y} plots the time evolution of the spin-$x$ and spin-$y$ populations for various values of $G$. At small $G$, clear oscillatory patterns are observed, with near-revivals of the initial spin-$y$ population. As $G$ increases, the irregular micro-oscillations become more prominent. Note that the spin-$z$ population, although not shown in Fig.~\ref{subfig:panel-y}, closely follows the spin-$y$ dynamics with a relative phase shift of $\pi/2$, consistent with the rotation symmetry of the model in the $y$--$z$ plane.

\subsubsection{Spin--boson entanglement}

A useful comparative reference is the well-known spin--boson model, which describes a two-level system coupled to an environment of harmonic oscillators. In that model, dephasing and loss of coherence are standard features and have been studied both analytically and numerically. For example, Orth \emph{et al.} \cite{PhysRevB.87.014305} studied nonperturbative decoherence in driven spin--boson systems, observing suppression of coherence and decay of the Bloch-vector magnitude as the coupling increases.

\begin{figure}[bpt]
    \centering
    \includegraphics[width=0.5\linewidth]{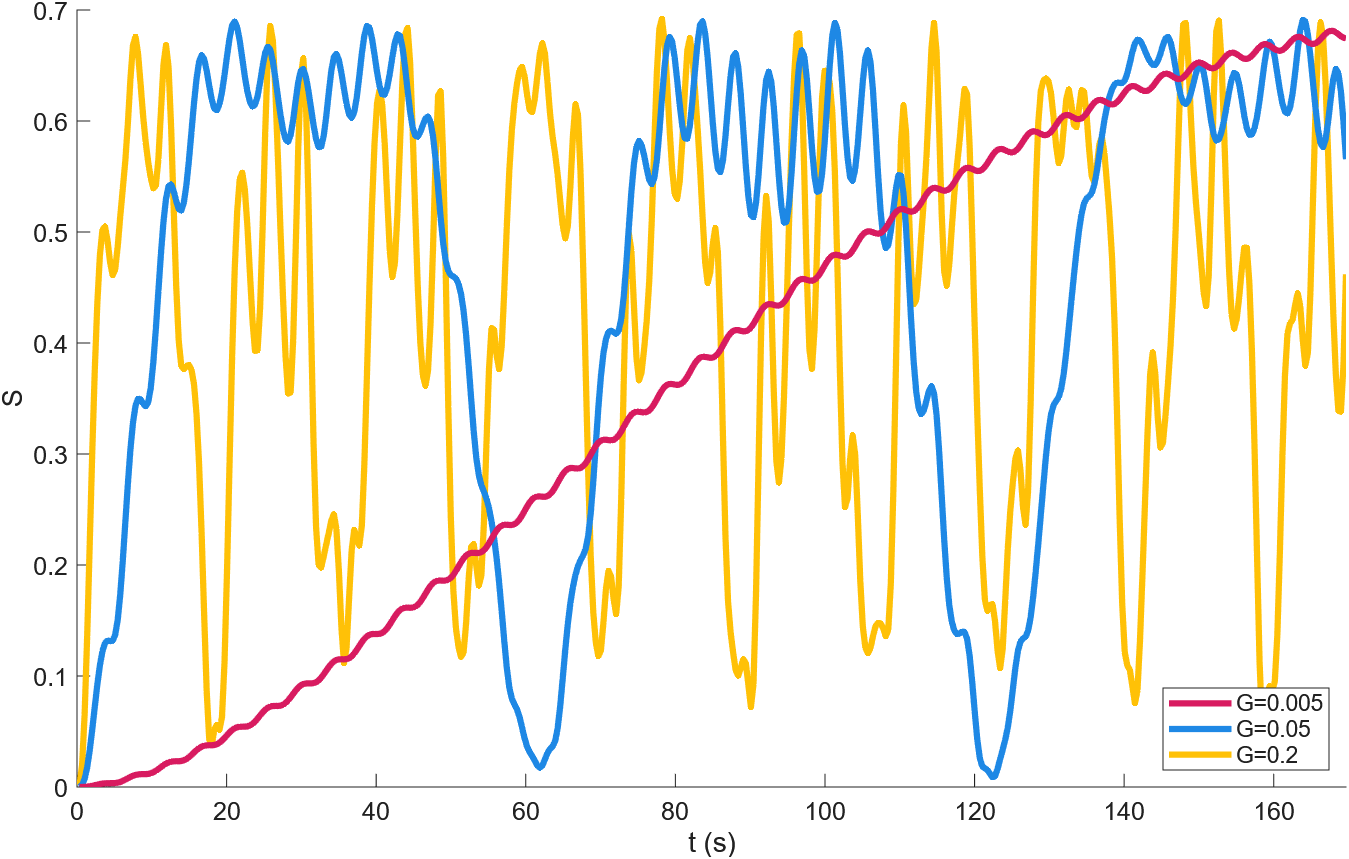}
    \caption{Entanglement entropy $S$ given in Eq.~\eqref{eq:vne} between the spin and bosonic subsystems, following the evolution generated by Hamiltonian~\eqref{eq:H-final} with $\mu=1$ and truncation at $N=14$. The spin is initially prepared in the $x$ direction.}
    \label{fig:vne}
\end{figure}

Since we limit the discussion to the small-coupling regime, where the Bloch vector does not shrink noticeably, we instead study numerically the decoherence of the spin density matrix due to the bosonic modes, i.e.\ the entanglement between them. Here we label the spin subsystem by $S$ and the bosonic modes by $E$ to emphasise the system--environment viewpoint. From the evolution of the reduced density matrix,
\begin{equation}
\rho_S(t) = \Tr_E\!\left(\ket{\psi(t)}\bra{\psi(t)}\right),
\end{equation}
we compute the time-dependent von Neumann entanglement entropy,
\begin{equation}\label{eq:vne}
    S(\rho_S) = - \sum_{i} \lambda_i \ln(\lambda_i) \qq{} \text{for } \lambda_i > 0 \text{ the positive eigenvalues of } \rho_S.
\end{equation}
The entanglement entropy $S(\rho_S)$ quantifies the entanglement between the bosonic subsystem and the spin subsystem. This is plotted in Fig.~\ref{fig:vne}, which shows that in the weak-coupling regime the entanglement is periodic. The figure considers the spin to be initialised along the $x$-axis and follows the same conventions as the previous figures: $\mu=1$ and the bosonic operators are truncated numerically at dimension $N=14$. The maximal entanglement entropy shown in the figure, $S\approx0.7$, corresponds to maximal entanglement for a two-level spin state, since the maximal entropy is given by $\ln(d)$, with $d$ the Hilbert-space dimension of $\rho_S$, i.e.\ $S_{\rm max}=\ln 2\approx 0.69$.

\vspace{5pt}
\subsection{Optical cavity quantum simulation}
\vspace{10pt}

We now present a cavity-QED implementation designed to simulate the specific minimal gravity--matter model described by Hamiltonian~\eqref{eq:H-final}. The possibility of tuning the couplings arbitrarily allows us to explore generic qualitative behaviours, such as coherence, entanglement, and decoherence of the spin caused by the quantum nature of gravity. The minimal model Hamiltonian in Eq.~\eqref{eq:H-final} describes a spin-$1/2$ particle interacting with two bosonic modes, which is well suited for experimental implementation using cavity quantum electrodynamics (cavity-QED) platforms. In particular, a single atom coupled to a high-finesse optical cavity supporting two orthogonal cavity modes offers a highly controllable realisation of such dynamics with long coherence times, suitable for the realisation of the desired time evolutions \cite{haroche2006exploring,RevModPhys.87.1379}.

In a typical cavity-QED setup, an individual atom is strongly coupled to the quantised electromagnetic field inside a high-finesse cavity, where coherent interactions dominate over dissipative processes. To encode the two bosonic degrees of freedom $\alpha$ and $\beta$ appearing in Eq.~\eqref{eq:H-final}, we can employ a bimodal cavity that simultaneously supports two nearly degenerate modes, distinguished, for example, by their polarisation or spatial structure. The atom itself provides the effective spin-$1/2$ system, with two long-lived hyperfine or Zeeman sublevels representing the $\ket{\uparrow}$ and $\ket{\downarrow}$ spin states.

The atom--cavity coupling can be engineered to produce terms proportional to $\sigma^x$ and $\sigma^y$ via appropriately chosen laser drives and cavity detunings. For example, in the strong-coupling regime, the cavity and atomic transitions hybridise to form an anharmonic ladder of dressed states, as described by the Jaynes--Cummings Hamiltonian \cite{PhysRevLett.118.133604}. This Hamiltonian naturally includes effective spin-boson coupling terms of the form $\left(\alpha + \alpha^\dagger\right)\sigma^y$ and $\left(\beta + \beta^\dagger\right)\sigma^x$, as required in Eq.~\eqref{eq:H-final}.

A concrete realisation can be achieved by simultaneously driving the atom and the cavity modes with appropriately tuned lasers. By choosing the laser amplitudes and detunings carefully, one can selectively couple each cavity mode to a specific atomic transition. One cavity mode may be chosen to drive spin rotations around the $x$-axis, while the other mediates rotations around the $y$-axis. The strength of these interactions is directly related to the effective spin-boson coupling parameter $g_{\text{s-b}}$, which in this setup can be tuned through the atom--cavity coupling strength, the detuning, and the drive amplitudes. Notably, such selective coupling strategies have been successfully employed to realise single- and two-photon blockades \cite{PhysRevLett.118.133604}, as well as atom--photon state mapping and deterministic single-photon generation \cite{Kuhn2015}.

The initialisation of the system into the state \eqref{eq:initial-state} can be performed by optically pumping the atom into a well-defined electronic state that encodes the spin and by cooling the cavity modes close to their vacuum state, which corresponds to the bosonic ground states $\ket{0}_\alpha \ket{0}_\beta$ in our model. The subsequent dynamical evolution of the spin can then be monitored by detecting the polarisation or temporal correlations of photons leaking from the cavity, providing direct access to the spin observables $\sigma_i(t)$ as well as the populations of the bosonic modes. State-of-the-art cavity-QED experiments already allow for the precise preparation, control, and readout of such hybrid atom--cavity systems with high fidelity and single-quantum resolution \cite{PhysRevLett.118.133604,Kuhn2015}. Therefore, our proposed minimal model can be directly implemented with existing optical cavity technology, paving the way for controlled laboratory simulations of spin dynamics in fluctuating spacetime backgrounds.

\vspace{5pt}
\section{Conclusions}\label{sec:conclusion}
\vspace{10pt}

We have introduced and analysed a minimal model for spin dynamics in a fluctuating quantum background, consisting of a single spin-$1/2$ system coupled to two self-interacting bosonic modes. This toy model captures essential features of matter interacting with a dynamical spacetime environment while remaining amenable to exact diagonalisation and controlled numerical study.

By preparing the spin in different initial orientations, we uncovered the population dynamics between the spin and bosonic modes and quantified their mutual entanglement. For weak coupling, the spin exhibits near-periodic oscillations and partial revivals, indicating reversible dynamics with limited backaction from the bosonic modes. As the coupling increases, these revivals appear to be gradually suppressed, but the minimal model of Eq.~\eqref{eq:H-final} eventually breaks down outside its regime of validity.

We also showed that this model is not purely theoretical: it can be directly realised in current cavity-QED platforms using a single atom coupled to a bimodal optical cavity. In such setups, the spin corresponds to atomic hyperfine states, while the bosonic modes are implemented, for example, by orthogonally polarised cavity photons with tunable self-interactions. The dynamical regimes discussed in this work are therefore accessible with existing experimental tools.

Our findings suggest that minimal quantum simulators can provide valuable insights into effective matter--gravity interactions in a fully quantum regime. Beyond the minimal instance, our full lattice model gives rise to gravitational fluctuations of a similar type to those appearing in the emergent gravitational theory that governs the low-energy behaviour of fractional quantum Hall liquids \cite{PhysRevX.7.041032,PhysRevLett.120.141601}, thus providing a controllable platform for exploring geometric responses and quantum analogues of curved spacetime in strongly correlated systems. Extensions of this model, such as extending it to $(3+1)$ dimensions, adding more bosonic modes, or including measurement backaction, could shed further light on how quantum information behaves in fluctuating or emergent spacetimes. Furthermore, while this model retains only the resonant modes, thus isolating the coherent and non-Markovian dynamics, considering dissipation from the coupling to a continuum and performing a full dissipative treatment is an interesting direction for future study.

\vspace{5pt}
\section*{Acknowledgments}
\vspace{10pt}

M.S. research projects are supported by the National Research Foundation, Singapore through the National Quantum Office, hosted in A*STAR, under its Centre for Quantum Technologies Funding Initiative (S24Q2d0009). M.S. research is supported by the Ministry of Education, Singapore under the Academic Research Fund Tier 1 (FY2022, A-8000988-00-00). P.S.-R. has been supported by a Young Scientist Training Program (YST) fellowship at the Asia Pacific Center for Theoretical Physics (APCTP) through the Science and Technology Promotion Fund and the Lottery Fund of the Korean Government. P.S.-R. has also been supported by the Korean local governments in Gyeongsangbuk-do Province and Pohang City. This work was partially supported by the EPSRC grants EP/Z533634/1 and UKRI1337:Anyons24.

\bibliographystyle{utphys}
\bibliography{Refs}

\appendix

\section{Jaynes--Cummings model approximation}\label{app:JC}

The Hamiltonian in Eq.~\eqref{eq:H-final} can partly be approximated by the Jaynes--Cummings model. Here we show that the coupling with the $\beta$ mode works via a spin-dependent displacement, while the coupling to the $\alpha$ mode works via excitation exchange by deriving the Jaynes--Cummings limit.

Rewriting the Hamiltonian in Eq.~\eqref{eq:H-final} in terms of $\sigma^\pm$, which act as ladder operators in the spin basis and are defined as $\sigma^y = \sigma^+ + \sigma^-$ and $i \sigma^z = \sigma^+ - \sigma^-$, gives
\begin{align}
    H = \sqrt{2} \sigma^x + \sqrt{2}(\alpha^\dagger \alpha + \beta^\dagger \beta) + g_\text{s-b} \left[ (\alpha^\dagger +\alpha)(\sigma^+ + \sigma^-) + (\beta^\dagger +\beta)\sigma^x \right]. \label{eq:JC1}
\end{align}
We then move to a displaced frame to remove the linear terms in the $\beta$ mode. The Hamiltonian becomes
\begin{align}
    \tilde{H} 
    &= D^\dagger H D 
    = \sqrt{2} \sigma^x + \sqrt{2}(\alpha^\dagger \alpha + \beta^\dagger \beta) + g_\text{s-b} (\alpha^\dagger +\alpha)(\sigma^+ e^{\sqrt{2}g_\text{s-b}} + \sigma^- e^{-\sqrt{2}g_\text{s-b}}) - \frac{g^2_\text{s-b}}{2}, \label{eq:JC2} \\
    &\text{with } D\left(-\frac{g\sigma^x}{\sqrt{2}}\right) = \exp\!\left[-\frac{g\sigma^x}{\sqrt{2}}(\beta^\dagger-\beta)\right].
\end{align}
As was shown in \cite{koch2023quantum}, the coupling between the spin and the $\beta$ mode causes a spin-dependent displacement of the $\beta$ mode. Switching to the interaction picture (rotating frame), denoted by the subscript $I$, we find
\begin{align}
    \tilde{H}_I(t) = g_\text{s-b} \left( \alpha\sigma^+ e^{i\sqrt{2}t} e^{\sqrt{2}g_\text{s-b}} + \alpha^\dagger\sigma^- e^{-i\sqrt{2}t} e^{-\sqrt{2}g_\text{s-b}} + \alpha^\dagger\sigma^+ e^{i3\sqrt{2}t} e^{\sqrt{2}g_\text{s-b}} + \alpha\sigma^- e^{-i3\sqrt{2}t} e^{-\sqrt{2}g_\text{s-b}}\right). \label{eq:JC3}
\end{align}
In Eq.~\eqref{eq:JC3}, the first two (co-rotating) terms show the exchange of population, while the latter two (counter-rotating) terms show either double excitation or double annihilation. Here, $\omega_0 = \sqrt{2}$ and the transition frequency is $\omega_T = 2\sqrt{2}$, which gives the detuning $\Delta\omega=\omega_T-\omega_0 = \sqrt{2}$. Thus, we are not exactly on resonance, but $\Delta \omega < \omega_T-\omega_0$. Since the counter-rotating terms oscillate three times faster than the co-rotating terms, and by setting $G$ such that $g_\text{s-b}\ll \omega_T-\omega_0$, we neglect the counter-rotating terms. Going back to the Schr\"odinger picture gives the Jaynes--Cummings limit of the displaced Hamiltonian:
\begin{equation}
    \tilde{H}^\text{JC} = \sqrt{2} \sigma^x + \sqrt{2}(\alpha^\dagger \alpha + \beta^\dagger \beta) + g_\text{s-b} \left(\alpha \sigma^+ e^{\sqrt{2}g_\text{s-b}} + \alpha^\dagger \sigma^- e^{-\sqrt{2}g_\text{s-b}} \right). \label{eq:JC4}
\end{equation}
In the Jaynes--Cummings limit, the total population is preserved; the spin and bosonic $\alpha$ mode exchange excitations, which can be seen, for example, in Fig.~\ref{subfig:2D}. Using the fact that the Jaynes--Cummings model preserves the total particle number, the average occupation number follows the standard two-level-system solution:
\begin{align}
    &\langle n_\alpha \rangle \sim \sin^2\left(\frac{\omega t}{2}\right), 
    &\langle \sigma^x \rangle \sim \cos^2\left(\frac{\omega t}{2}\right),
\end{align}
with $\omega$ the Rabi frequency. Reference \cite{koch2023quantum} found the population of the displaced $\beta$ mode to be
\begin{equation}
    \langle n_\beta(t) \rangle = 2 g^2 \sin^2\left(\frac{\sqrt{2}t}{2}\right),\label{eq:nbeta}
\end{equation}
where we have substituted $\omega_0 =\sqrt{2}$. The frequency matching of the two bosonic modes is not immediately apparent from these expressions, likely because Ref.~\cite{koch2023quantum}, from which Eq.~\eqref{eq:nbeta} was derived, does not include any spin interaction and instead treats it as a constant.

\end{document}